\documentclass[12pt]{article}
\usepackage{amsmath}
\usepackage{graphicx}
\usepackage{amsfonts}
\usepackage{amssymb}

\begin{document}




\strut\hfill AEI-2001-036 
\vfill
\begin{center}
{\large \bf OPEN MEMBRANES, p-BRANES AND \\[.5em] NONCOMMUTATIVITY OF
BOUNDARY STRING \\[.5em] COORDINATES} \vskip 2em
\vspace{1cm}
{\large {\sc Ashok Das$^\dagger$,
J. Maharana$^\ddagger$ and  A. Melikyan$^\dagger$ }}
\\[.5cm]
$^\dagger$ {\it Department of Physics and Astronomy\\
University of Rochester,
Rochester, NY 14627-0171, USA}
\\[0.3cm]
$^\ddagger${\it Max-Planck-Institut f\"ur Gravitationsphysik\\
Albert-Einstein-Institut, Am M\"uhlenberg 1
D-14476 Golm, Germany} \\[.5em] and\\[.5em]
{\it Institute of Physics\\ 
Bhubaneswar -751005, India}
\par
\end{center}\par
\vskip .5cm
\begin{center} {\bf Abstract} \end{center} 
We study the dynamics of an open membrane with a cylindrical topology, in
 the background of a constant  three form, whose boundary is attached to
 p-branes. The boundary closed string is coupled to a two form potential
 to ensure gauge invariance. We use the action, due to Bergshoeff, London 
 and Townsend, to study the noncommutativity properties of the boundary 
 string  coordinates. The constrained Hamiltonian formalism due to Dirac  
 is used to derive the  noncommutativity of coordinates. The chain of
 constraints is found to be finite for a suitable gauge choice, unlike the case
 of the static gauge, where the chain has an infinite sequence of terms.
 It is conjectured that the formulation of closed string field theory  
 may necessitate introduction of a star product which is both 
 noncommutative and nonassociative.

\vfill\eject

\section{Introduction:}

Recently, the study of noncommutative geometry, from the perspective
of string theory,
has attracted considerable attention. The noncommutativity of the
target space coordinates becomes manifest when a constant background
NS two form potential is 
introduced along the D-brane \cite{noncom1}. In the presence of the two-form
potential, the end points of the open strings attached to the D-brane
do not commute. The intimate connection between string theory, noncommutative
geometry and noncommutative Yang-Mills theories  
has been investigated from diverse points of view by Seiberg and Witten
in their seminal paper \cite{sw}. It is natural, therefore, to examine the
corresponding issue when an open membrane ends on a D-brane under
an analogous situation.
In this case, one would envisage a D-brane in flat space in the
background of a three-form tensor potential, with components tangential to
brane coordinates, since a membrane couples to the third rank
antisymmetric tensor just as a string couples to the  two-form potential.  

The motivation to study this problem arises from the conjectures that 
membranes provide a clue to understanding M-theory. The five perturbatively
consistent string theories are believed to be different phases of the
underlying M-theory and it is argued that the low energy limit of M-theory
is described by the $D=11$ supergravity theory, just as low energy limits of 
various super string theories go over to the corresponding supergravity theories
in ten dimensions. It is also well known that membrane and five brane 
appear naturally in eleven dimensional supergravity and already, there
have  been attempts
to study the $M5-$brane in the background of a constant $3$-form \cite{sw}.
Another motivation to study properties of open membranes arises from the
perspective of OM theory \cite{gop}. It has been conjectured that
five dimensional noncommutative open string theory on an $M5$ brane,
in the strong coupling limit, decouples from gravity, for a critical
value of  the field strength, and becomes equivalent to a theory of
light open membranes (OM theory decouples from gravity with a near
critical $3$-form field strength.).
Furthermore, some progress has been made in examining
how noncommutativity may arise by using open membranes
as probes on branes \cite{membr1,membr2}. 

It is worth while to mention, at this point, that the study of
noncommutativity properties of open membranes ending on branes
involves some subtleties  
compared to the case where one considers an open
string ending on D-branes. The action, in the case of string
theory, is chosen to be of Polyakov type, whereas for membranes the
conventional
choice has been the Nambu-Goto action, with additional terms which are
introduced from various other considerations. Furthermore, the
equations of  motion for a Nambu-Goto action are nonlinear, in contrast
to  the linear equations arising in the case of the Polyakov
formulation of the string theory. Therefore, one has to resort to some
approximation scheme while investigating the noncommutativity aspects
of  such a theory. We mention {\it en passant} that one of the easiest
ways  to bring out the
noncommutativity for an open string-D-brane system, in the presence of a
NS  B-field, is to scale the metric and the B-field suitably and then
take  $\alpha '\rightarrow 0$ limit. However, in the case of a 
membrane-brane  system, there is no such
obvious limit, as has been emphasized by several authors \cite{membr1,membr2}.

In this paper, we have investigated the noncommutativity property of
an open membrane-brane configuration from a different perspective.
In the first place, in order to circumvent some of the technical 
difficulties encountered in the Nambu-Goto action, we adopt a modified
version of the action due to Bergshoeff, London and
Townsend \cite{blt}, where one introduces a two-form gauge potential in the
world volume and where the tension is given the status of a dynamical variable 
(depends on world volume coordinates). The equations of motion derived from
the BLT action coincide with those coming from the Nambu-Goto action, 
when  one substitutes the solutions of the auxiliary equations for the
world volume two-form potential as well as the
tension, into the equation of motion for the  membrane coordinates. In
fact,  one can write down an additional contribution due to the space-time
dependent three form background coupled to the membrane and study the
symmetries associated with such a system \cite{jmward}. As we shall show, there
are  certain advantages
in adopting the BLT action. When an open string end is attached to the D-brane,
one introduces the coupling of the open string to a gauge field background
and the gauge invariance of the world sheet action implies that the field
strength associated with the gauge potential be constant for constant B-field.
Note that in the analogous  situation, where the open membrane (more
details about the membrane configuration later, which we basically
choose to have a cylindrical topology) ends on the brane, the boundary
is  that of
a closed string and, therefore, a gauge symmetry demands that one needs to
introduce a two-form B-field coupled to the boundary string just as a
boundary $D0$-brane 
couples to a gauge potential. Thus, we are set out to study noncommutativity
properties of the ends of the membrane which are coordinates of a string,
depending on world volume time and one of its spatial coordinates.

The  constrained Hamiltonian formalism due to Dirac \cite{dirac} is one 
of the most elegant and powerful techniques in which one can exhibit
this  feature of the
theory. Let us recapitulate how noncommutativity arises in an open
string theory when a constant B-field
is introduced along the brane direction. In the absence of B, the
coordinates  along the Dp-brane satisfy Neumann boundary condition,
$\partial_{\sigma}X^{\mu}=0, \mu =0,1..p$, at the boundary 
$\sigma =0, \pi$; here we take the target space metric to be the flat
Minkowski  metric, 
$\eta _{\mu\nu}$. When B is introduced, the boundary condition changes to
a mixed condition, namely, $\eta _{\mu\nu}\partial _{\sigma}X ^{\nu} +
B_{\mu\nu} \partial _{\tau} X^{\nu} =0$ at the boundaries. 
It has been suggested that the boundary condition can be used as a
primary constraint after eliminating $\dot{ X }^{\mu}$ in favor of
the  canonical momenta of the
string coordinates. One can then use the procedure of Dirac to derive
all  the secondary constraints of the theory,
identify the second class constraints and finally evaluate the Dirac brackets
between various phase space variables \cite{bdnoncom} to see
noncommutativity of coordinates.  

It is important to note that supersymmetry is not an essential ingredient in
the study of noncommutativity in the  geometry of a brane in
a constant  anti-symmetric field background.
Therefore, we shall consider a bosonic open membrane ending on a
brane.  However,
it is also important to keep in mind that this is a simplified model
and within  the
context of M-theory, it will be essential to consider super membranes. In
that frame work, one should consider the action in the background of the 
massless fields of eleven dimensional supergravity such as the
graviton and the  3-form
tensor. The gauge invariance will be lost if we do not introduce the 2-form
B-field for the open membrane, as alluded to above. When one introduces
open super membranes, all the supersymmetries are broken, even in the flat
Minkowski space. However, in the presence of topological defects as 
backgrounds \cite{susybr}, it is possible to construct supersymmetric
actions for an open  
supermembrane. These defects have a natural interpretation analogous
to the  end of
the world 9-plane in the Horava-Witten construction \cite{hw}. Thus,
keeping  in 
mind that one is likely to deal with supermembrane theories, we consider
an open  membrane ending on a p-brane with constant target space
metric and  three form potential along with a  2-form B-field coupled to the
closed  string on the boundary. 

Our approach is similar to that of Kawamoto and Sasakura. However,
with the modified form of the action, we are able to make some head
way with the 
computation of the matrix of constraints as well as the evaluation of
the  Dirac brackets (DB) in
a systematic manner, without linearizing the action as was
done at the outset in \cite{membr2}. It is worthwhile to
point  out one interesting feature that arises in the computation of
the  secondary constraints for the problem under consideration. To
start  with, in the Dirac formalism,
one identifies the primary constraints and requires that the canonical
Poisson bracket (PB) of these constraints with the total Hamiltonian must
vanish. In other words, the constraints must be stationary under time
evolution. As a consequence, either one generates new constraints, which
in turn generate further constraints and this goes on {\it ad
infinitum}, or  that this process
terminates in the sense that one does not generate any new constraint after
a finite number of iterations. However, as we will show, in the case of
the membrane, unlike in string theories, the velocities,
$\dot{X}^{\mu}$ cannot be written in terms of the canonical momenta
and, consequently, the boundary conditions cannot be written as
primary constraints on the phase space,
{\it unless one chooses a gauge to begin with}. In ref \cite{membr2}, the
Dirac analysis was carried out in the static gauge (in addition to the
linearized approximation for the action) and an infinite chain of
constraints was obtained.
In fact, a similar situation also arises in the case of open strings ending on 
D-branes. Namely, one knows that incorporation of boundary conditions
as primary constraints, in general, leads to an infinite chain of
constraints.  Surprisingly, however, we find that with an alternate,
suitable  choice
of gauge (one is free to choose a gauge), the constraint chain for the
same membrane system terminates. In other words, in this alternate
gauge,  after a finite number of iterations, new constraints are not
generated. On the other hand, as a consistency check,
when we do resort to the static gauge, we find an infinite 
chain of constraints, similar to those of ref \cite{membr2}; the two
chains do not coincide since the starting point of Hamiltonian analysis
are different (namely, the starting actions are different), due to
different gauge choices. It is well known
that noncommutativity depends on the choice of gauge (which, however,
cannot  be eliminated). That the nature of the chain of constraints
also depends on the choice of gauge is something that we had not seen
earlier.   

The paper is organized as follows. In section {\bf 2}, we first consider the 
Nambu-Goto action for the membrane and discuss some of its salient
features. The
resulting equations of motion are presented along with the relevant boundary 
conditions. Subsequently, the alternate action due to Bergshoeff, London
and Townsend \cite{blt} is introduced; it is recalled how the original
equations of motion are recovered from this action. In  section {\bf 3}, we 
proceed systematically with the Hamiltonian analysis to identify the 
primary constraints. We show how a gauge choice becomes essential to
carry out the Dirac procedure and choose a  suitable gauge in order to
facilitate the analysis of constraints. In section {\bf 4}, we carry out
the  analysis of constraints in some detail and show that, in this
particular choice of gauge, the chain of constraints terminates. In
section {\bf 5}, the evaluation of the Dirac brackets is discussed,
where we explicitly determine the Dirac brackets to linear order in
the background anti-symmetric field and demonstrate
how noncommutativity of the coordinates on the boundary arises.
We summarize our results in section {\bf 6} and end with
conclusions. In appendix {\bf 1}, we briefly indicate how the static
gauge leads to an infinite chain of constraints and in appendix {\bf
2}, we point out the essential structure of the Dirac brackets to quadratic
order in the anti-symmetric background field.

\section{The Action:}

It is well known that the action for a membrane can be described by a
Nambu-Goto action, much like the action for a string. Thus, for
example, the world volume action for an open membrane in
$11$-dimensions, interacting
with an anti-symmetric background field, $C_{MNP}$ can be described by
the  action
\begin{equation}
S=T\left( \int_{\Sigma_{3}}d^{3}\xi\left(\sqrt{g}-\frac{1}{6}\varepsilon
^{ijk}\partial_{i}X^{M}\partial_{j}X^{N}\partial_{k}X^{P}C_{MNP}\right)
+\int_{\partial\Sigma_{3}}B\right)\label{1}
\end{equation}
where $T$ represents the tension of the membrane.
Here $M,N,P=0,1,..,10$ are indices of the $11$-dimensional
target space, $\xi=(\tau,\sigma_{1},\sigma_{2})$ are the coordinates of
the world volume of the membrane with the corresponding indices taking
values $i,j,k = 0,1,2$, $g=\det g_{ij}$, where
$g_{ij}= G_{MN}\partial_{i}X^{M}\partial_{j}X^{N};$  is the  induced
metric on the membrane (We use a metric with signature $(+,-,-)$ on
the world volume and one with signature $(+,-,-,\cdots ,-)$ in the
target space.). In
addition to the anti-symmetric tensor
background on the world volume, here we also have a boundary term
which is required for gauge invariance of the action. Namely, with the
boundary term, the action is invariant under gauge 
transformations of the kind %
\begin{align}
C  &  \rightarrow C+d\Lambda\nonumber\\
& \nonumber\\
B  &  \rightarrow B-\Lambda\label{2}
\end{align}
where $\Lambda$ is the local, two form parameter of
transformation. Combining  $C$ and $dB$ into a single 3-form
\begin{equation}
A=C+dB \label{3}
\end{equation}
we can rewrite the action (\ref{1}) in the following form:
\begin{equation}
S=T\,\int_{\Sigma_{3}}d^{3}\xi\left(\sqrt{g}-\frac{1}{6}\varepsilon
^{ijk}\partial_{i}X^{M}\partial_{j}X^{N}\partial_{k}X^{P}A_{MNP}\right)
\label{4}
\end{equation}
For simplicity, we are going to choose both $G_{MN}$ and $A_{MNP}$ to
be constants (In fact, we will choose $G_{MN}=\eta_{MN}$ from now
on.).  In such a
case,  it is well known that this action
can describe an open membrane ending on p-branes in the eleven
dimensional target space. 

In fact, let us note that the action (\ref{4}) leads to the equations
of motion
\begin{equation}
\partial_{i}\left(\sqrt{g}g^{ij}\eta_{MN}\partial_{j}X^{N} - \frac{1}{2}
\varepsilon^{ijk}A_{MNP}\partial_{j}X^{N}\partial_{k}X^{P}\right) = 0\label{4'}
\end{equation}
along with the boundary conditions
\begin{equation}
n_{i}\left(\sqrt{g}g^{ij}\eta_{MN}\partial_{j}X^{N} - \frac{1}{2}
\varepsilon^{ijk}A_{MNP}\partial_{j}X^{N}\partial_{k}X^{P}\right)\delta
X^{M} 
= 0\label{4''}
\end{equation}
where $n_{i}$ represents the unit normal vector. If we assume that our
membrane has a cylindrical topology
characterized by $0\leq \sigma_{1}\leq \pi, 0\leq \sigma_{2}\leq 2\pi$
in units of some length scale, this can, in fact, describe an open
membrane terminating on p-branes. The Hamiltonian analysis, following
from the action  (\ref{4}), has been carried out in \cite{membr2} with
the gauge choices (for reparameterization invariance)
\begin{eqnarray}
X^{0} & = & \tau,\qquad\;\;\;\; \tau\in (\infty,\infty)\nonumber\\
X^{9} & = & \sigma_{1}L,\qquad \sigma_{1}\in [0,\pi]\nonumber\\
X^{10} & = & \sigma_{2}R,\qquad \sigma_{2}\in [0,2\pi)\label{config}
\end{eqnarray}
where the radius of the compactified direction, $X^{10}$, is $R$. 
(As we will see later, in the case of membranes, gauge fixing becomes
essential before carrying out an analysis of constraints.) 
It has been shown, in such a case, how  noncommutativity arises in
the Dirac brackets. However, because of the
nonlinear nature of such a formulation, the analysis in \cite{membr2} was
done only in the linearized approximation (as well as other
approximations) and it would be nice to see if one can do the analysis
without restricting to the linearized approximation.

An alternate description for the membrane is through the first order
action of the form (due to Bergshoeff, London and Townsend) \cite{blt}
\begin{equation}
S_{BLT}=\int_{\Sigma_{3}}d^{3}\xi\,\frac{1}{2V}(g-\widetilde{\mathcal{F}}^
{2})
\label{5}
\end{equation}
Here we have defined
\begin{equation}
\widetilde{\mathcal{F}}\equiv
\varepsilon^{ijk}\widetilde{\mathcal{F}}_{ijk} =\varepsilon^{ijk}
(F_{ijk}+\frac{1}{6}A_{ijk})
\label{6}%
\end{equation}%
where
\begin{eqnarray}
F_{ijk} & = &\partial_{\lbrack i}U_{jk]} = \partial_{i}U_{jk} + {\rm
cyclic} \label{7}\\
A_{ijk} & = &
\partial_{i}X^{M}\partial_{j}X^{N}\partial_{k}X^{P}A_{MNP}\label{8}%
\end{eqnarray}
with $A_{MNP}$ defined earlier. Clearly, $V(\xi)$ and $U_{ij}(\xi)$ are
auxiliary field variables. If we now look at the equations of motion
following from the action (\ref{5}), we obtain
\begin{eqnarray}
\partial_{i}\left(\frac{1}{V}gg^{ij}\partial_{j}X^{N} \eta_{MN}\right. &-&\left.\frac{1}{2}
\frac{\widetilde{\mathcal{F}}}{V}\varepsilon^{ijk}\partial_{j}X^{N}
\partial_{k}X^{P}A_{MNP}\right)=0 \label{10}\\
\partial_{i}\left(\frac{\widetilde{\mathcal{F}}}{V}\right)&=& 0 \label{11}\\
g &=& \widetilde{\mathcal{F}}^{2} \label{12}
\end{eqnarray}
The constraint equations in (\ref{11}) and (\ref{12}) can be easily
solved. Choosing the solutions as
\begin{equation}
\widetilde{\mathcal{F}} = TV = \sqrt{g}\label{13}
\end{equation}
with $T$ representing the tension in Eq. (\ref{4}) and substituting
these into Eq. (\ref{10}), we see that the
dynamical equation for the coordinates coincides with the one in (\ref{4'})
following from the action in (\ref{4}).

Similarly, we note that the boundary conditions following from the
action (\ref{5}) take the forms
\begin{eqnarray}
n_{i}\left((\frac{1}{V})gg^{ij}\partial_{j}X^{N}\eta
_{MN}\right.&-&\left.(\frac{\widetilde{\mathcal F}}{2V})\varepsilon^{ijk}\partial_{j}X^{N}\partial
_{k}X^{P}A_{MNP}\right)\delta X^{M}=0 \label{14}\\
\varepsilon^{ijk}n_{i}\delta U_{jk}&=& 0 \label{15}
\end{eqnarray}
Using the solutions (\ref{13}) for the auxiliary field
equations, we see that Eq. (\ref{14}) reproduces the boundary
condition (\ref{4''}) following from the action (\ref{4}). The second
boundary condition,  (\ref{15}), can simply be satisfied by choosing
$U_{01}$ to be a constant along $\sigma_{1}.$

Due to the cylindrical topology of the membrane, in the presence of
p-branes, the boundary
condition  (\ref{4''}) or equivalently (\ref{14}) reduces to
\begin{equation}
\sqrt{g}g^{1j}\partial_{j}X_{\mu}-\frac{1}{2}\varepsilon^{1jk}
\partial_{j}X^{\nu
}\partial_{k}X^{\rho}A_{\mu\nu\rho}\mid_{\sigma_{1}=0,\pi}=0,\text{
\ \ \ \ \ \ \ }\mu,\nu=0,1,...,p\text{\ \ \ \ } \label{16}
\end{equation}
\begin{equation}
X^{a}=x_{0}^{a},\text{ \ \ \ \ \ \ \ \ \ \ \ \ \ }a=p+1,...,10 \label{17}
\end{equation}
where the coordinates, $x_{0}^{a}$, specify the  positions of the
p-branes  at the two boundaries $\sigma_{1}=0,\pi$.
\bigskip

\section{Gauge fixing and the Hamiltonian:}

We take the first order action (\ref{5}) as the starting point of our
canonical description of the system. The goal, of course, is to find
the Hamiltonian, implement the boundary conditions
as primary constraints, determine all the constraints of the theory
and  derive the Dirac brackets for the system. The
problem, however, is that the action for the membrane is highly
constrained. Of course, even string theory has constraints. However,
as we will show next, the constraints in the case of membranes are
much more difficult to deal with and necessitate  gauge fixing. 

To see the difficulty, let us note that the Lagrangian density of the
action (\ref{5}) is singular in the sense that the determinant of the
coefficient matrix
multiplying the quadratic terms in velocities (namely, the Hessian matrix)
\[
\frac{\partial^{2}L_{BLT}}{\partial\overset{.}{X}^{M}\partial\overset{.}{X}^{N}}
\]
where\qquad\qquad\
\begin{equation}
L_{BLT}=\frac{1}{2V}(g-\widetilde{\mathcal{F}}^{2}) \label{19}
\end{equation}
vanishes. This, of course, reflects the fact that there are
constraints in the system. To see the nature of the constraints in a
simple manner, let us introduce the following notations. Let us
separate the world volume indices into time and space as $i = (0,a)$
where $a=1,2$. In that case, it is easy to check that the determinant
of the metric can be written in the factorized form
\begin{equation}
g = \det g_{ij} = \left(g_{00}-g_{0a}\overline{g}^{ab}g_{b0}\right)
\det g_{ab}\label{20}
\end{equation}
where, as we have defined earlier,
\[
g_{ab} = \partial_{a}X^{M}\partial_{b}X^{N}\eta_{MN},\qquad a,b=1,2
\]
and $\overline{g}^{ab} (\neq g^{ab})$ is the inverse of this in the
two dimensional subspace. With this notation, it is clear that the
momentum, conjugate to $X^{M}$, is given by
\begin{eqnarray*}
P_{M} & = & \frac{\partial L_{BLT}}{\partial \dot{X}^{M}}\\
 & = & \frac{1}{V}\left(\left(\dot{X}_{M} -
 \partial_{a}X_{M}\overline{g}^{ab}g_{b0}\right) \det g_{ab} - \frac{1}{2}
 \widetilde{\mathcal{F}}\varepsilon^{ab}\partial_{a}X^{N}\partial_{b}X^{P}
A_{MNP}\right)
\end{eqnarray*}
so that we can write
\begin{equation}
\left(\eta_{MN} -
\partial_{a}X_{M}\overline{g}^{ab}\partial_{b}X_{N}\right)\dot{X}^{N}
= \frac{1}{\overline{g}}\left(V P_{M} +
\frac{1}{2}\widetilde{\mathcal{F}}\varepsilon^{ab}\partial_{a}X^{N}
\partial_{b}X^{P}A_{MNP}\right)\label{21}
\end{equation}
where we have introduced the notation $\overline{g}=\det
g_{ab}$. There are several things to note from the structure of
Eq. (\ref{21}). First, it follows immediately from this equation as
well as the definitions of $g_{ab}$, that
\begin{equation}
P_{M}\partial_{a}X^{M} = 0,\qquad\qquad a=1,2\label{22}
\end{equation}
These are, in fact, the two generators of reparameterization symmetry
along the $\sigma_{1},\sigma_{2}$ directions. The analogous relation
in string theory is
\[
P_{M}\partial_{\sigma}X^{M} = 0
\]
where$P_{M}\partial_{\sigma}X^{M}$  corresponds to the generator of
$\sigma$  reparameterization invariance. The generator of
reparameterization along the $\tau$ direction is the Hamiltonian as we
will see later. However, more important from our point of view is the
fact that the matrix multiplying $\dot{X}^{N}$, on the left hand side
of Eq. (\ref{21}), is a transverse projection operator and, therefore,
the velocities cannot be expressed in terms of the phase space
variables. Since the boundary conditions in Eq. (\ref{16}) involve
velocities, this also means that, as it stands, the boundary
conditions cannot be written as primary conditions on the phase space
so that the Dirac analysis cannot be carried out in a conventional
manner. We would like to emphasize that this is a new feature that is
not  present in the Hamiltonian analysis of strings. It is not hard to
see that the origin of this difficulty lies in the reparameterization
invariance in our theory. If we fix a gauge, then the velocities can,
in fact, be expressed in terms of phase space variables and the Dirac
analysis can be carried out. It is well known that local
symmetries, such as 
reparameterization invariance, lead to first class
constraints which need to be gauge fixed. However, normally, this is done
after the constraint analysis has been completed and all the
constraints have been classified into first class and second class
constraints.  In the present
case, on the other hand, we cannot even carry out the constraint analysis,
because the boundary conditions cannot be written as phase space
constraints, {\it unless we fix a gauge}.

It is clear, therefore, that to carry out the Hamiltonian analysis for
the membrane, it is necessary to start with a gauge fixed action instead
of the action in (\ref{5}). A conventional gauge choice (static gauge)
such as
\[
X^{0} = \tau,\quad g_{0a} = 0\quad a=1,2
\]
would fix all the reparameterization invariances of the theory and
would  allow the
velocities to be inverted. In this case, one can carry out the
constraint analysis, which leads to the expected behavior that the
boundary conditions induce an infinite chain of constraints (We will
describe this briefly in appendix {\bf 1}.). However, the infinite number of
such constraints are highly nonlinear and, therefore, are not readily
amenable to calculating Dirac brackets.

We will, therefore, choose an alternate gauge condition
which brings out an interesting feature of our analysis, namely, that
with a  suitable gauge choice, the chain of constraints can terminate.
A finite set of constraints is clearly much easier to handle, in calculating
Dirac brackets. Let us
look at the action (\ref{5}) in the gauge
\begin{equation}
g_{0a} = 0,\qquad a=1,2\label{22'}
\end{equation}
Namely, we are going to fix only the reparameterization invariance
along the spatial directions (This is equivalent to fixing two of the
spatial coordinates which we take to be normal to the brane.). 

In this gauge,  action (\ref{5}) takes the form
\begin{equation}
S = \int_{\Sigma_{3}} d^{3}\xi\,\frac{1}{2V} \left(\overline{g}
\dot{X}^{M}\dot{X}_{M} -
\widetilde{\mathcal{F}}^{2}\right)\label{22''}
\end{equation}
This action still has a $\tau$ reparameterization invariance, but, as
we will see, it does not interfere with the Hamiltonian analysis of
the system. It is now straightforward to determine the canonical
momenta from Eq. (\ref{22''}) and they take the forms
\begin{eqnarray}
P_{M}  &  = &\frac{\partial L}{\partial
\dot{X}^{M}}=\frac{\overline{g}}{V}\dot{X}_{M}-\frac{\widetilde
{\mathcal{F}}}{2V}\varepsilon^{0ab}\partial_{a}X^{N}\partial_{b}X^{P}
A_{MNP}\nonumber\\
\Pi^{(U)ab}  & = & \frac{\partial L}{\partial \dot{U}_{ab}} = -
\frac{3\widetilde{\mathcal{F}}}{V}\varepsilon^{oab} = -
\frac{3\widetilde{\mathcal{F}}}{V}\varepsilon^{ab}
\nonumber\\
\Pi^{(U)oa} & = &\frac{\partial L}{\partial \dot{U}_{0a}}  \approx 0\nonumber\\
P_{V}  &  = & \frac{\partial L}{\partial \dot{V}} \approx 0 \label{23}
\end{eqnarray}
Here $P_{M},\Pi^{(U) ab},\Pi^{(U)oa},$ and $P_{V}$ are momenta conjugate to
the fields $X^{M},U_{ab}$, $U_{0a}$ and $V$ respectively and it is clear
that we have two primary constraints in the theory given by the
last two equations in (\ref{23}).

It is also clear from Eq. (\ref{23}) that the velocities can be
inverted and we have
\begin{equation}
\dot{X}_{M}  =\frac{V}{\overline{g}}\mathcal{P}_{M}\label{24}
\end{equation}
where we have defined
\begin{equation}
\mathcal{P}_{M}  \mathcal{\equiv}P_{M}-\frac{1}{6}\Pi^{(U)ab}\partial
_{a}X^{N}\partial_{b}X^{P}A_{MNP} \label{24'}
\end{equation}
The canonical Hamiltonian can now be determined and has the form
\begin{equation}
H_{can}=P^{M}\dot{X}_{M}+\Pi^{(U) ab}\dot{U}_{ab}-L \label{25}
\end{equation}
and, using (\ref{23}) and (\ref{24}), can be written in the form
\begin{equation}
H_{can}=\frac{V}{2\overline{g}}\mathcal{P}^{2}-\frac{V}{72}(\Pi^{(U)ab}
\varepsilon_{ab})^{2}+2\Pi^{(U)ab}\partial_{a}U_{0b} \label{26}
\end{equation}
We can now derive the boundary conditions following from the action
(\ref{22''}) which, using Eq. (\ref{24}), can be converted to phase
space constraints of the form
\begin{equation}
\left.\left(\overline{g}^{a1}\partial_{a}X_{\mu}\mathcal{P}^{2}+\frac{1}{3}\Pi
^{(U)a1}\mathcal{P}^{\nu}\partial_{a}X^{\lambda}A_{\mu\nu\lambda}\right)
\right|_{\sigma_{1}=0,\pi}\approx0 \label{27}
\end{equation}
With this, we are now ready to carry out the constraint analysis for the
system, which we take up in the next section. 

We note here that in the constraint analysis of ref \cite{membr2},
the three form background
was chosen to be purely magnetic, namely, $A_{0IJ}=0, 
 I,J =1,2,...,p$ and only the spatial components, $A_{IJK}$, are nonzero.
However, our constraint analysis in the next section and the evaluation of
the Dirac brackets in section {\bf 5} do not crucially depend on such a choice
for the A-field. On the other hand, let us note that if $A_{0IJ}$ were
nonzero, then it is likely that the DB of the time component of the
boundary  string coordinate 
$X^0 (\tau, \sigma _2)$ with a spatial component of string coordinate
will be nonzero and the same result will hold for the DB of time-time
coordinates. Although, the consequences of noncommutativity of these
coordinates have not been investigated in detail, it is well known from
the analysis of the corresponding D-brane open string that 
such noncommutativity of space-time
and time-time coordinates pose some difficulty in formulating the 
perturbative Feynman rules \cite{feyn}. On the other hand, there are
persuasive reasons to believe that space and time coordinates may not
commute, based on some general arguments as well as 
certain string and M-theoretic analysis leading to space-time uncertainty 
relations \cite{thft, yon,lennys}. In view of all these possibilities,
we have kept our analysis quite general without choosing any specific
form for the $A$-field . 

\section{Constraint analysis:}

In the previous section, we determined the canonical Hamiltonian of
the system in the gauge (\ref{22'}) and noted the two primary
constraints following  from the
definition of the conjugate momenta. Adding to these the boundary
constraints, the complete set of primary constraints can be written as
\begin{eqnarray}
\varphi_{1}  & = & P_{V}\approx 0\nonumber\\
\varphi_{2}^{a}  & = & \Pi^{(U)oa}\approx 0\nonumber\\
\varphi_{3\mu}  & = & (\overline{g}^{a1}\partial_{a}X_{\mu}\mathcal{P}
^{2}+\frac{1}{3}\Pi^{(U)a1}\mathcal{P}^{\nu}\partial_{a}X^{\lambda}A_{\mu\nu\lambda 
})\delta(\sigma_{1})\approx 0\nonumber\\
\varphi_{4\mu}  & = & (\overline{g}^{a1}\partial_{a}X_{\mu}\mathcal{P}
^{2}+\frac{1}{3}\Pi^{(U)a1}\mathcal{P}^{\nu}\partial_{a}X^{\lambda}A_{\mu\nu\lambda 
})\delta(\sigma_{1}-\pi)\approx 0 \label{28}
\end{eqnarray}
Adding these primary constraints to the canonical Hamiltonian (\ref{26}), we
obtain the Hamiltonian for the system to be
\begin{equation}
H=\frac{V}{2\overline{g}}\mathcal{P}^{2}-\frac{V}{72}(\Pi^{(U)ab}
\varepsilon_{ab})^{2}+2\Pi^{(U)ab}\partial_{a}U_{0b}+c\varphi_{1}+
k_{a}\varphi_{2}^{a}+\lambda^{\mu}\varphi_{3\mu}+\widetilde{\lambda}^{\mu
}\varphi_{4\mu} \label{29}
\end{equation}
where $c,k_{a},\lambda^{\mu},$ and $\widetilde{\lambda}^{\mu}$ are
Lagrange multipliers. 

The analysis for the consistency of constraints can now be carried out
in a straightforward manner. The consistency condition
\begin{equation}
\dot{\varphi}_{1} = \left\{\varphi_{1},\int H\right\} \approx 0
\end{equation}
leads to
\begin{eqnarray}
\lambda^{\mu} & = & \widetilde{\lambda}^{\mu} = 0\nonumber\\
\varphi_{5} & = & \frac{{\mathcal{P}}^{2}}{2\overline{g}} -
\frac{1}{72} \left(\Pi^{(U)ab}\varepsilon_{ab}\right)^{2} \approx
0\label{30}
\end{eqnarray}
There are a couple of points  to be  noted from this.
First, the vanishing
of the Lagrange multipliers, $\lambda^{\mu}$ and
$\widetilde{\lambda}^{\mu}$, can be understood intuitively as
follows. We have already seen in Eq. (\ref{24}) that the velocities
can be expressed in terms of the phase space variables as
\[
\dot{X}_{M}=\frac{V}{\overline{g}}{\mathcal{P}}_{M}
\]
On the other hand, if we calculate directly, using the Hamiltonian
(\ref{29}), we have
\begin{equation}
\dot{X}_{M}=\left\{X_{M}, \int H\right\} \label{31}
\end{equation}
and we find that the two are compatible only if
$\lambda^{\mu} =0 = \widetilde{\lambda}^{\mu}$. 
The  second point to note
from Eq. (\ref{30}) is that the non-evolution of the constraint
$\varphi_{1}$ leads to a secondary constraint, $\varphi_{5}$, which,
as we  will see shortly,
is simply the constraint reflecting the $\tau$ reparameterization
invariance of the theory. But, for the present, we only note that the
consistency of this secondary constraint leads to no further constraint.

Let us next note that the consistency of the second constraint in
Eq. (\ref{28}),
\[
\dot{\varphi}^{a}_{2} = \left\{\varphi^{a}_{2}, \int H\right\} \approx
0
\]
leads to a secondary constraint of the form
\begin{equation}
\varphi^{a}_{6}   = \partial_{b}\Pi^{(U)ba} \approx 0 \label{32}
\end{equation}
This is the analog of Gauss' law in electrodynamics  and it can be
easily checked that the consistency of this constraint leads to no
new constraints. Furthermore, it is clear now that with
Eq. (\ref{32}), the constraint in Eq. (\ref{30}) simply says that the
canonical Hamiltonian (and, therefore, $H$) vanishes in this theory,
which is a reflection of the $\tau$ reparameterization invariance of the
theory. Notice that, since $U_{ab}$ is a cyclic variable in our
theory, it follows that 
$\Pi^{(U)ab}$ is a constant of motion. Thus, together with
Eq. (\ref{32}), this implies that $\Pi^{(U)ab}$ is truly a constant, which
in turn implies, from Eq. (\ref{30}), that
$\frac{\mathcal{P}^{2}}{\overline{g}}$ is a constant. Without loss of
generality we choose  
\begin{equation}
\frac{\mathcal{P}^{2}}{\overline{g}}=1\label{33}
\end{equation}

With the help of Eq. (\ref{33}), the boundary conditions can now be
rewritten (reduced) as
\begin{eqnarray}
\varphi_{3\mu} & = &
(\overline{g}\,\overline{g}^{a1}\partial_{a}X_{\mu}+ \frac{1}{3}
\Pi^{(U)a1}\mathcal{P}^{\nu}\partial_{a}X^{\lambda}A_{\mu\nu\lambda})\delta
(\sigma_{1})\approx 0\nonumber\\
\varphi_{4\mu} & = &
(\overline{g}\,\overline{g}^{a1}\partial_{a}X_{\mu}+\frac{1}{3}
\Pi^{(U)a1}\mathcal{P}^{\nu}\partial_{a}X^{\lambda}A_{\mu\nu\lambda})\delta
(\sigma_{1}-\pi)\approx0\label{34}
\end{eqnarray}
Consistency of the boundary constraint $\varphi_{3\mu}\approx 0$,
\[
\dot{\varphi}_{3\mu}=\left\{\varphi_{3\mu}, \int H\right\} \approx 0
\]
leads to the secondary constraint,
\begin{eqnarray}
\varphi_{7\mu} & = & \delta(\sigma_{1})\left[\overline{g}^{a1}\partial
_{a}[\frac{V}{\overline{g}}\mathcal{P}_{\mu}]+\varepsilon^{ab}\partial
_{a}X_{\mu}\partial_{(b}X^{\lambda}\partial_{2)}[\frac{V}{\overline{g}
}\mathcal{P}_{\lambda}]+\frac{V}{3\overline{g}}\Pi^{(U)a1}A_{\mu\nu\lambda
}\mathcal{P}^{\nu}\partial_{a}\mathcal{P}^{\lambda}\right.\nonumber\\
 & &
\left. -\frac{1}{3}\Pi^{(U)a1}A_{\mu\nu\lambda}\partial_{a}X^{\lambda}
\partial
_{c}[V\overline{g}^{bc}\partial_{b}X^{\nu}]\right]\label{36}
\end{eqnarray}
In deriving Eq. (\ref{36}), we have used the relations
$\left\{\mathcal{P}_{\mu}(\sigma),\mathcal{P}_{\nu}(\sigma')\right\} 
\approx  0$, which follows from the  Gauss' law constraint, as well as
the  following PB 
\begin{eqnarray}
\left\{{\mathcal P}_{\mu}{\mathcal
P}^{\mu}(\sigma),\frac{1}{\overline{g}(\sigma')}\right\} &
 = & -
\frac{4}{\overline{g}(\sigma')}\overline{g}^{ab}(\sigma')\partial_{a}X_{\nu}
(\sigma')\partial_{b}\delta
(\sigma-\sigma^{^{\prime}})\label{37}
\end{eqnarray}

Since the Hamiltonian (\ref{29}) contains a term of the form $c P_{V}$ and the
secondary constraint $\varphi_{7\mu}$ depends on $V$ as well as
$\partial V$, consistency of this new constraint
\[
\dot{\varphi}_{7\mu} = \left\{\varphi_{7\mu}, \int H\right\} \approx 0
\]
simply determines the
Lagrange multiplier $c$ and leads to no further constraint. The
expression for
the Lagrange multiplier is complicated and  its explicit form is not very
crucial for our analysis; therefore, we do not present it here.   An identical
analysis goes through for the constraint
$\varphi_{4\mu}$ at the other boundary and generates only one
secondary  constraint
$\varphi_{8\mu}$, whose structure is identical to that of
$\varphi_{7\mu}$ except that it is at the other boundary. What we have
found  is truly remarkable. We may recall that the boundary constraints, in
the context of string theory as well as in the analysis of   
 constraints in \cite{membr2}, led to an infinite chain of
constraints. In contrast, we find that in a particular gauge, the boundary
constraints for the case at hand lead  
 to only one secondary constraint at each boundary. Namely, the
chain of constraints actually terminates which is quite desirable from
the point of view of calculating Dirac brackets. We have also carried out the 
constraint analysis, starting from our action in the static gauge and
indeed, we find that the infinite chain of constraints do appear. That
is, had we worked in a different gauge, there would be
an infinite chain of constraints as normally expected. It is known in
the  literature that
the noncommutativity arising in Dirac brackets depends on the gauge
choice (which, however, cannot be eliminated). However, what we find
here is that the nature of the constraint chain itself depends on the
choice of gauge. This is, in fact, the only example that we are aware
of, where the constraint chain for boundary conditions terminates
\cite{shahin}.

\section{Dirac brackets:}

Since we have determined all the constraints of our theory, it is now
straightforward, in principle, to determine the Dirac
brackets. However, we note that the boundary constraints, (\ref{27})
and (\ref{36}),
are, in
particular, highly nonlinear and, consequently, evaluation of the
inverse of the matrix of constraints is, in general,  a very difficult
problem. Things, however, do simplify enormously if we use a weak field
approximation for $A_{\mu\nu\lambda}$. In this case, it is easy to
determine the inverse of the matrix of constraints in an iterative manner
to any order in the $A_{\mu\nu\lambda}$ field. Let us demonstrate this
by first calculating the Dirac bracket to linear order in
$A_{\mu\nu\lambda}$. In appendix {\bf 2}, we will indicate the structure
of the Dirac brackets to second order in this field.

Let us note that among the entire set of constraints including the
primary constraints
$\varphi_{1},\varphi^{a}_{2},\varphi_{3\mu},\varphi_{4\mu}$ and the
secondary constraints
$\varphi_{5},\varphi^{a}_{6},\varphi_{7\mu},\varphi_{8\mu}$, only the
boundary 
constraints $\varphi_{3\mu},\varphi_{7\mu},\varphi_{4\mu}$ and
$\varphi_{8\mu}$ are second class constraints. The other constraints
are all first class and can be handled by choosing appropriate gauge
fixing conditions. These (first class) constraints do not influence
the evaluation of the Dirac bracket of the coordinates
$\{X^{\mu},X^{\nu}\}_{D}$ (in which we are interested) and, consequently,
we will ignore them for our analysis. Furthermore, the analysis of the
Dirac bracket using the constraints at the second boundary
($\sigma_{1}=\pi$) is completely parallel to that at the first
boundary ($\sigma_{1}=0$) so that we will describe the analysis using
only one set of constraints, say $\varphi_{3\mu},\varphi_{7\mu}$. 

The constraints $\varphi_{3\mu},\varphi_{7\mu}$ are second class and,
therefore, the Dirac bracket between the coordinates takes the form
\begin{eqnarray}
\{X_{\mu}(\sigma),X_{\nu}(\sigma^{\prime})\}_{D} & = & -\int
d^{2}\sigma'' d^{2}\sigma'''\,\{X_{\mu}(\sigma
),\phi_{A}(\sigma'')\}C^{-1}{}^{AB}(\sigma'',\sigma''')\nonumber\\
 &  & \qquad \times \{\phi_{B}(\sigma'''),X_{\nu}(\sigma')\}
\label{DB1}
\end{eqnarray}
where $\phi_{A}\equiv(\varphi_{3\mu},\varphi_{7\mu})$ and
\begin{equation}
C_{AB}(\sigma,\sigma')=\left(
\begin{array}
[c]{cc}%
\{\varphi_{3\mu}(\sigma),\varphi_{3\nu}(\sigma')\} &
\{\varphi_{3\mu}(\sigma),\varphi_{7\nu}(\sigma')\}\\
\{\varphi_{7\nu}(\sigma'),\varphi_{3\mu}(\sigma)\} &
\{\varphi_{7\mu}(\sigma), \varphi_{7\nu}(\sigma')\}
\end{array}
\right)  \label{C}
\end{equation}

We can, of course, calculate exactly all the brackets entering into
the matrix, $C_{AB}$.  However, determining the inverse matrix exactly
is a technically nontrivial problem. It is here that the weak field
approximation is of immense help (We want to emphasize that there is
no other approximation used in our derivations.). To proceed, let us
note that
\begin{eqnarray}
\{\varphi_{3\mu}(\sigma),X_{\lambda}(\sigma^{\prime})\} & = & -\frac{\delta
(\sigma_{1})}{3}\Pi^{(U)a1}\partial_{a}X^{\rho} A_{\mu\lambda\rho
}\delta(\sigma-\sigma')\label{fi3X}\\
\{\varphi_{7\mu}(\sigma),X_{\lambda}(\sigma')\}  & = & \delta(\sigma
_{1})\left(\varepsilon^{ab}g_{b2}\partial_{a}\left(\frac{V}{\overline{g}}\delta
(\sigma-\sigma')\right)\eta_{\mu\lambda}\right.\nonumber\\
 &  & \quad +\varepsilon^{ab}\partial_{a}X_{\mu
}\partial_{(b}X_{\lambda}\partial_{2)}\left(\frac{V}{\overline{g}}\delta
(\sigma-\sigma')\right)\nonumber\\
 & & 
\left.+\frac{V}{3\overline{g}}\Pi^{(U)a1}A_{\mu\lambda\rho}\left((\partial
_{a}\mathcal{P}^{\rho})-\mathcal{P}^{\rho}
\partial_{a}\right)\delta (\sigma-\sigma')\!\!\right) \label{fi4x}
\end{eqnarray}
Here and in what follows $[a_{1}\cdots a_{n}]$ would stand for
anti-symmetrization,
while $(a_{1}\cdots a_{n})$ would denote symmetrization of indices. We
also note that,
unless explicitly denoted, all quantities depend on $\sigma$
($\sigma'$ dependence will be explicitly displayed). Let us
parameterize  these relations as
\begin{eqnarray}
\{\varphi_{3\mu}(\sigma),X_{\lambda}(\sigma')\} & = & S_{\mu\lambda}
\delta(\sigma-\sigma') \label{U}\\
\{\varphi_{7\mu}(\sigma),X_{\lambda}(\sigma')\} & = & T_{\mu\lambda}
\delta(\sigma-\sigma')+U_{\mu\lambda}\delta(\sigma-\sigma')
\label{V}
\end{eqnarray}
where
\begin{eqnarray}
S_{\mu\lambda}(\sigma) & = & -\frac{\delta(\sigma_{1})}{3}\Pi^{(U)a1}\partial
_{a}X^{\rho} A_{\mu\lambda\rho} \label{DefU}\\
T_{\mu\lambda} (\sigma) & = &
\delta(\sigma_{1})\Big(\varepsilon^{ab}g_{b2}
\partial_{a}\left(\frac{V}{\overline{g}}\right)
\eta_{\mu\lambda}\nonumber\\
 &  & \qquad\quad
+\varepsilon^{ab}\partial_{a}X_{\mu}\partial_{(b}X_{\lambda}
\partial_{2)}\left(\frac{V}{\overline{g}}\right)
\label{DefV}\\
 & &
+\frac{V}{\overline{g}}\left(\varepsilon^{ab}g_{b2}\eta_{\mu\lambda} 
+\varepsilon^{bc}\partial_{b}X_{\mu}\partial_{c}X_{\lambda}\delta_{2}^{a}
\right.\nonumber\\
 & &\qquad \left. - \varepsilon^{ab}
\partial_{b}X_{\mu}\partial_{2}X_{\lambda}\right)\partial_{a}\Big)\nonumber\\
 &\equiv& \Omega_{\mu\lambda}(\sigma) +
\Gamma_{\mu\lambda}^{a}(\sigma)\partial_{a} \nonumber\\
U_{\mu\lambda} (\sigma) & = & \delta(\sigma_{1}) \frac{V}
{3\overline{g}}\,\Pi^{(U) a1}A_{\mu\lambda\rho}\left((\partial_{a}{\mathcal
P}^{\rho}) - \mathcal{P}^{\rho}\partial_{a}\right)
\label{DefW}%
\end{eqnarray}
Here, we have separated the Poisson bracket structures into terms that
depend on $A_{\mu\nu\lambda}$ and those which do not. Thus, for example
$S_{\mu\nu}$ and $U_{\mu\nu}$ are linearly dependent on
$A_{\mu\nu\lambda}$ while $T_{\mu\nu}$ is independent of
$A_{\mu\nu\lambda}$. This is, in fact, quite natural and useful since
we are going to
be expanding in powers of $A_{\mu\nu\lambda}$. Furthermore, note that
it is  best to think of
$S,T,U$ as operators acting on the delta function, so that $S$ is a
multiplicative operator while $T$ and $U$ each has a
multiplicative part as well as a part that is linear in the derivative
operator. For example, $\Omega_{\mu\lambda}$
represents the multiplicative operator in $T_{\mu\lambda}$, while
$\Gamma_{\mu\lambda}^{a}\partial_{a}$ corresponds to the terms with
the derivative operator. 

Let us next analyze the structure of the inverse of the matrix 
$C_{AB}$. If we represent  
the matrix obtained from the mutual PB of second class constraints in 
the generic form as 
\begin{equation}
C_{AB} = \left(\begin{array}{cc}
                a & b\\
                c & d
                \end{array}\right)
\end{equation}
where $a,b,c,d$ are themselves matrices (with indices suppressed for
simplicity),
then, it can be easily checked that the inverse has the form
\begin{equation}
(C^{-1})^{AB} = \left(\begin{array}{cc}
                       \alpha & \beta\\
                       \gamma & \delta
                       \end{array}\right)
\end{equation}
where
\begin{eqnarray}
\alpha & = & (a - bd^{-1}c)^{-1}\nonumber\\
\beta  & = & (c - db^{-1}a)^{-1}\nonumber\\
\gamma & = & (b - ac^{-1}d)^{-1}\nonumber\\
\delta & = & - b^{-1}a (c - db^{-1}a)^{-1}\label{inv}
\end{eqnarray}
Once we know the constraint matrix (namely, $a,b,c,d$), it is clear
that we can calculate the inverse to any given order in
$A_{\mu\nu\lambda}$ by a suitable expansion technique, 
using Eq. (\ref{inv}). The Dirac
bracket for the coordinates, of course, has the (symbolic) form
\begin{equation}
\left\{X,X\right\}_{D} = \left(\widetilde{S}\alpha S +
(\widetilde{T}+\widetilde{U})\gamma S + \widetilde{S}\beta (T+U) +
(\widetilde{T}+\widetilde{U})\delta (T+U)\right)\label{DB}
\end{equation}
where $\widetilde{S},\cdots $ represent the transposed operators
corresponding to $S,\cdots $ in the complete space.

Let us note next that we are interested in terms which are linear in
$A_{\mu\nu\lambda}$ in the Dirac bracket in Eq. (\ref{DB}). Since
$S$ and $U$ are already linear in this field, the Dirac bracket
simplifies and takes the form
\begin{equation}
\left\{X,X\right\}_{D} = \left(\widetilde{T}\gamma^{(0)} S +
\widetilde{S}\beta^{(0)} T +
\widetilde{T}\delta^{(0)} U + \widetilde{U}\delta^{(0)} T +
\widetilde{T}\delta^{(1)} T\right)\label{dirac}
\end{equation}
where, clearly, $\beta^{(0)},\gamma^{(0)},\delta^{(0)}$, in the first
four  terms on the right hand side, do not contain  
$A_{\mu\nu\lambda}$, while
$\delta^{(1)}$ in  the last term is linear in this field (The
superscript signifies the number of $A_{\mu\nu\lambda}$ contained.).

To evaluate the Dirac bracket, let us note that
\begin{eqnarray}
\left\{\phi_{3\mu}(\sigma),\phi_{3\nu}(\sigma')\right\} & = &
\frac{\delta (\sigma _{1})\delta (\sigma _{1}^{^{\prime
}})}{3}\Pi^{(U) a1}\Big(
-\varepsilon ^{bc}\partial _{a}X^{\lambda}
(\sigma^{\prime })g_{2c}A_{\mu\nu \lambda }\partial _{b}\delta (\sigma -\sigma 
^{\prime })\nonumber\\
 & & + \varepsilon ^{bc}\partial_{a}X^{\lambda
}g_{2c}(\sigma ^{\prime })A_{\mu\nu \lambda}\partial_{b}\delta (\sigma
-\sigma ^{\prime })\nonumber\\
 & & + \varepsilon ^{bc}\partial
_{a}X^{\rho }
(\sigma^{\prime })\partial _{b}X_{\mu }A_{\nu \lambda \rho }(\partial
_{(c}X^{\lambda }\partial _{2)}\delta (\sigma -\sigma ^{\prime }))\nonumber\\
 & &  - \varepsilon
^{bc}\partial_{a}X^{\rho }
\partial_{b}X_{\nu }(\sigma ^{\prime })A_{\mu \lambda \rho }(\partial
_{(c}^{^{\prime }}X^{\lambda }(\sigma ^{\prime })\partial _{2)}\delta (\sigma
-\sigma ^{\prime }))\nonumber\\
 & &  +\frac{3}{\overline{g}}\Pi ^{(U) b1}A_{\mu \lambda \rho }A_{\nu
\alpha \gamma }\eta ^{\alpha \rho }\mathcal{P}^{\lambda}\partial _{a}X^{\gamma
}(\sigma ^{\prime })\partial _{b}\delta (\sigma -\sigma ^{\prime }) 
\nonumber\\
 & & -\frac{3}{\overline{g}}\Pi ^{(U) b1}A_{\mu \lambda \rho }A_{\nu
\alpha \gamma }\eta ^{\lambda \gamma }\mathcal{P}^{\alpha }(\sigma ^{\prime
})\partial _{a}X^{\rho }\partial _{b}^{^{\prime }}\delta (\sigma -\sigma
^{\prime })\Big)\label{a}
\end{eqnarray}
It is clear from this that the element \lq\lq $a$'' in the coefficient
matrix $C_{AB}$ is linearly and quadratically dependent on
$A_{\mu\nu\lambda}$. Since we are interested only in terms which are linear in
$A_{\mu\nu\lambda}$ in the Dirac bracket, we can neglect the terms
quadratic in $A_{\mu\nu\lambda}$ in this Poisson bracket to write
\begin{equation}
\left\{\phi_{3\mu}(\sigma),\phi_{3\nu}(\sigma')\right\} =
V_{\mu\nu}(\sigma,\sigma') + W_{\mu\nu}(\sigma,\sigma')
\end{equation}
where $V_{\mu\nu},W_{\mu\nu}$ are respectively the symmetric and the
anti-symmetric parts (in the Lorentz index) of the terms in Eq. (\ref{a})
which are linear in $A_{\mu\nu\lambda}$, namely,
\begin{eqnarray}
V_{\mu\nu} (\sigma,\sigma') & = &
\frac{\delta (\sigma _{1})\delta (\sigma _{1}')}{6}\Pi^{(U)
a1}\varepsilon^{bc} \Big(
\partial_{a}X^{\rho}(\sigma')\partial _{b}X_{(\mu }A_{\nu )\lambda
\rho } \partial_{(c}X^{\lambda }\partial _{2)} \nonumber\\
 &  & \quad -\partial _{a}X^{\rho }\partial_{b}X_{(\nu }(\sigma')
A_{\mu )\lambda \rho }\partial_{(c}^{^{\prime }}X^{\lambda }(\sigma')
\partial _{2)}\Big)\delta (\sigma -\sigma')
\end{eqnarray}
and
\begin{eqnarray}
W_{\mu\nu} (\sigma,\sigma') & = &
\frac{\delta (\sigma _{1})\delta (\sigma _{1}')}{6}
\Pi ^{(U)a1}\varepsilon ^{bc}\Big(-2\partial _{a}X^{\lambda}(\sigma')
g_{2c}A_{\mu \nu \lambda}\partial _{b}\nonumber\\
 &  & + 2\partial _{a}X^{\lambda}
g_{2c}(\sigma')A_{\mu \nu \lambda}\partial _{b}
+ \partial _{a}X^{\rho}(\sigma')\partial _{b}X_{[\mu }
A_{\nu]\lambda \rho}\partial_{(c}X^{\lambda}\partial _{2)}\nonumber\\
 &  &
- \partial_{a}X^{\rho}\partial_{b}X_{[\nu }(\sigma')
A_{\mu ]\lambda\rho}\partial_{(c}^{^{\prime }}X^{\lambda}(\sigma')
\partial _{2)}\Big)\delta (\sigma -\sigma')
\end{eqnarray}
Thus, we see that since \lq\lq $a$'' is already linear in the field
$A_{\mu\nu\lambda}$, it now follows from Eq. (\ref{inv}) that
\begin{eqnarray}
\beta^{(0)} & = & \left. c^{-1}\right|_{A_{\mu\nu\lambda}=0}\nonumber\\
\gamma^{(0)} & = & \left. b^{-1}\right|_{A_{\mu\nu\lambda}=0}\nonumber\\
\delta^{(0)} & = & 0\nonumber\\
\delta^{(1)} & = & -\left. b^{-1}ac^{-1}\right|\label{inv1}
\end{eqnarray}
where in the last expression $b,c$ are restricted to be evaluated with
$A_{\mu\nu\lambda}=0$. It is worth noting from (\ref{inv1}) and
(\ref{DB}) that since $\delta^{(0)}=0$, the Dirac bracket vanishes to
zeroth order in the $A_{\mu\nu\lambda}$ field. This is consistent with
our understanding that noncommutativity arises in the presence of an
anti-symmetric background.

Since we already know the structure of the element $a$, it is clear
that, at this order, all the relevant elements of the inverse matrix
are determined
completely from a knowledge of the elements $b,c$ which are
related to each other. Let us next analyze the structure of these
elements. Defining,
\begin{equation}
M_{\mu\nu} (\sigma,\sigma') =
\left.\left\{\phi_{3\mu}(\sigma),\phi_{7\nu}(\sigma')\right\}
\right|_{A_{\mu\nu\lambda}=0}
\end{equation}
it is straightforward to calculate and show that  
\begin{eqnarray}
M_{\mu\nu}(\sigma,\sigma')  & = &-\delta (\sigma _{1})\delta
(\sigma _{1}')\varepsilon ^{ab}\varepsilon ^{cd}\left[g_{b2}(\sigma') 
\partial _{a}^{^{\prime }}\left( \frac{V(\sigma ^{^{\prime }})}
{\overline{g}(\sigma ^{^{\prime }})}\partial _{(d}^{^{\prime }}\delta
(\sigma -\sigma')\partial _{2)}X_{\nu }\partial
_{c}X_{\mu}\right.\right.\nonumber\\
& &\qquad\qquad\qquad +\left. \frac{V(\sigma')}{\overline{g}(\sigma
^{^{\prime }})}
\partial _{c}^{^{\prime }}\delta (\sigma -\sigma ^{\prime })g_{d2}\eta
_ {\mu\nu }\right)  \nonumber\\
& & - \partial _{a}X_{\nu }(\sigma')\partial_{(b}X_{\lambda }(\sigma')
\partial _{2)}^{^{\prime }}\left(
\frac{V(\sigma')}{\overline{g}(\sigma')}
\partial_{(d}^{^{\prime }}\delta (\sigma -\sigma')\partial _{2)}
X^{\lambda}\partial _{c}X_{\mu }\right.\nonumber\\
& &\qquad\qquad +\left.\left. \frac{V(\sigma')}{\overline{g}(\sigma')}
\partial _{c}^{^{\prime }}\delta (\sigma -\sigma')g_{d2}
\delta_{\mu }^{\lambda }\right)\right]\nonumber\\
&=&G_{\mu \nu }(\sigma ,\sigma')+F_{\mu \nu }(\sigma ,\sigma')\label{77}
\end{eqnarray}
where $G_{\mu\nu}$ and $F_{\mu\nu}$ represent the symmetric and the
antisymmetric parts of $M_{\mu\nu}$ in the Lorentz indices. More explicitly,
\begin{eqnarray}
G_{\mu\nu}(\sigma,\sigma')  & = &\!\!\! -\frac{\delta (\sigma_{1})\delta
(\sigma _{1}')}{2}\varepsilon
^{ab}\varepsilon^{cd}\left[g_{d2}(\sigma')
\partial _{a}^{^{\prime }}\left(
\frac{V(\sigma')}{\overline{g}(\sigma')}
\partial_{(d}^{^{\prime }}\delta (\sigma -\sigma')
\partial _{2)}X_{(\nu}\partial _{c}X_{\mu )}\right)\right.\nonumber\\
&& + g_{b2}(\sigma')\partial _{a}^{^{\prime}}\left(\frac
{V(\sigma')}{\overline{g}(\sigma')}
\partial _{c}^{^{\prime }}\delta (\sigma -\sigma')g_{d2}\eta _{\mu\nu
}\right)\nonumber\\
&& + \partial _{a}X_{(\nu }(\sigma ^{\prime})
\partial _{(2}X_{\lambda }(\sigma')\partial _{b)}^{^{\prime}}
\left(\frac{V(\sigma')}{\overline{g}(\sigma')}
\partial _{(d}^{^{\prime }}\delta (\sigma -\sigma')
\partial_{2)}X^{\lambda }\partial _{c}X_{\mu )}\right)\nonumber\\
&& + \partial _{a}X_{(\nu }(\sigma')\partial _{(2}X_{\mu )}(\sigma')
\partial _{b)}^{^{\prime}}\left(\partial _{c}^{^{\prime }}
\delta (\sigma -\sigma')g_{d2}\right)\Big]
\label{99}\\
F_{\mu\nu}(\sigma,\sigma^{^{\prime}}) & = &\!\!\!
-\frac{\delta (\sigma_{1})\delta (\sigma _{1}')}{2}\varepsilon ^{ab}
\varepsilon^{cd}\Big[g_{b2}(\sigma')\partial _{a}^{^{\prime }}
\left( \frac{V(\sigma')}{\overline{g}(\sigma')}
\partial_{(d}^{^{\prime}}\delta (\sigma -\sigma')
\partial _{2)}X_{[\nu}\partial _{c}X_{\mu ]}\right)\nonumber\\
&& + \partial _{a}X_{[\nu }(\sigma')
\partial _{(b}X_{\lambda}(\sigma')
{\partial^{^{\prime}}}_{2)}\left( \frac
{V(\sigma')}{\overline{g}(\sigma')}\partial_{(d}^{^{\prime }}
\delta (\sigma -\sigma')\partial _{2)}X^{\lambda}
\partial _{c}X_{\mu ]}\right)\nonumber\\
&& + \partial _{a}X_{[\nu }(\sigma')
\partial _{(b}X_{\mu ]}(\sigma'){\partial^{^{\prime}}}_{2)}
\left( \frac{V(\sigma')}{\overline{g}(\sigma')}
\partial _{c}^{^{\prime }}\delta(\sigma -\sigma')g_{d2}\right)\Big]
 \label{88}
\end{eqnarray}
Thus, we  can write the inverse of $M_{\mu\nu}$ formally as
\begin{equation}
(M^{-1})^{\mu\nu}(\sigma,\sigma')=\left(\left(G+F\right)^{-1}
\right)^{\mu\nu}(\sigma,\sigma') \label{minverse}
\end{equation}
The important thing to note here is that, since neither
$G_{\mu\nu}$ nor $F_{\mu\nu}$ vanishes, the
inverse exists and contains both symmetric and antisymmetric
parts. This, therefore, determines $b^{-1}$ which is related to
$c^{-1}$ through a negative sign. Therefore, we are now ready to write
down the Dirac bracket between the coordinates in the linear
approximation in the $A_{\mu\nu\lambda}$ field.

Substituting all of this into Eq. (\ref{dirac}), we find that
the Dirac bracket has the form
\begin{eqnarray}
\{X_{\mu}(\sigma),X_{\nu}(\sigma')\}_{D}  & = &
\left[\widetilde{T}_{\mu\lambda} 
\left(\left(G+F\right)^{-1}\right)^{\lambda\rho}S_{\rho\nu}\right.\nonumber\\
  &  &   - \widetilde{S}_{\mu\lambda}\left(\left(\left(G+F\right)^{-1}\right)^
{\lambda\rho}T_{\rho\nu}\right)\label{final}\\ 
 &  &
+\left.\widetilde{T}_{\mu\lambda}\left(\left((F+G)^{-1}(V+W)(F+G)^{-1}\right)^
{\lambda\rho}T_{\rho\nu}\right)\!\right]\!(\sigma,\sigma')\nonumber
\end{eqnarray}
where the derivative operators in the factor $T$ on the right are
assumed to act on the factor to the left within the
parenthesis with a negative sign (We will give the explicit form of
this Dirac bracket in appendix {\bf 2}). It is worth emphasizing here
that because of the structure of the boundary constraints in
Eqs. (\ref{34}) and (\ref{36}), the $\sigma_{1}$ coordinate is fixed
at the boundary (to be $0,\pi$) and, therefore, the Dirac bracket,
evaluated at equal $\tau$, effectively depends only on the world
volume coordinates $\tau,\sigma_{2}$. This shows that the boundary
string coordinates indeed become
noncommutative  in the presence of an anti-symmetric background field and what
is even more interesting is that they have a structure that is quite
analogous to that in the case of strings.

\section{Summary and Discussions}

We have studied an open membrane, with cylindrical geometry, ending on
p-branes.
The boundary of the open membrane on the brane is a closed string. We
have confined our attention to a background configuration where the target
space metric and the three-form potential are constant. We have also
incorporated a two-form potential, on the boundary, whose presence is
necessary to maintain gauge invariance.

The world volume action is conventionally chosen to be of Nambu-Goto form. We
have adopted a modified action which has some distinct advantages as discussed
in the text. We have treated the boundary conditions as primary
constraints  and have
shown that, one can carry out the Dirac formalism without restricting
to the linearized approximation of the action. We have also 
introduced a gauge choice, different from the one adopted in ref \cite{membr2},
and have shown that the Dirac procedure, in this gauge, leads to a
finite number constraints. As a 
consequence, we are able to compute the PB brackets of
all  the second class constraints which are necessary
for the evaluation of Dirac brackets.

Since the second class constraints are highly nonlinear, one, however,
has to adopt an approximation scheme in order to determine the inverse
of the matrix of constraints. In the absence of a length scale
(i.e. there is no analogue of $\alpha '$
here), we evaluate the inverse matrix as well as the Dirac brackets
order by order in $A_{\mu\nu\lambda}$, assuming
that this constant background can be taken to be small (we have chosen 
background metric to be $\eta _{\mu\nu}$ for simplicity). As a consistency check, 
we find that the DB between coordinates on the boundary (circle) vanishes when
the $A$-field is set to zero. We have computed the noncommutativity (strictly
speaking DB) up to quadratic terms in the three form field. Since, we carry
out the computations in a different gauge, our DB relations differ from ref.
\cite{membr2} (As is well known, the noncommutativity in the Dirac
brackets is gauge dependent.).

We would like to note here that we have also carried out the double
dimensional reduction \cite{mduff} (although we do not discuss it
here) as another consistency
check, where one of the coordinates say $X^{10}$ is compactified on a circle
and it is chosen to be identical to one of the world volume coordinates,
say $\sigma _2$. Furthermore, as is customary for double dimensional reduction
on a circle, if we assume that the rest of the coordinates are independent of
$\sigma _2$, then we obtain the reduced action corresponding to 
that of Cederwall and Townsend \cite{ct} as was employed in 
\cite{jmward}. Thus, if one had started from  the action \cite{ct},
one could have carried out the constraint analysis through a slightly 
different route for open strings ending on D-branes.

We may recall, that the two form potential, introduced on the boundary 
to ensure gauge invariance, is actually space-time dependent for a
constant field strength ($H=dB$ in form notation). Recently, it has been
pointed out, in the context of strings ending on D-branes, that
such space-time dependent (even for constant H-field) theories result in
noncommutative field theories which do not respect associativity \cite{spn}.
The $\star$ product, in such a case, is replaced by a generalized
product  constructed by  Kontsevich \cite{kt}.
Note that the boundary of the membrane is a closed string, in contrast
to the case where the boundary of an open string ending on a D-brane is
a point, a
$D0$-brane. When this $D0$-brane couples to a gauge potential, whose
field strength is taken to be  constant
for a constant B-field, the gauge theory becomes noncommutative. As
is  argued in \cite{spn}, for a space-time dependent B-field, in the case of
open  strings ending on a D-brane, on the other hand, one has a
noncommutative and nonassociative $\bullet$ product.  
In our study of open membranes with a cylindrical topology,  we have
closed  strings on the boundary. With a space-time dependent B-field
whose field strength is space-time dependent, therefore, we speculate that
the underlying field theory of strings will be described by an
underlying geometry which is noncommutative as well as nonassociative.
At this stage, we do not have any insight to comment on how noncommutativity
of string coordinates, in general, will modify the formulation of
open string field theory \cite{witten} or closed string field 
theory \cite{barton}. However, based on the developments in noncommutative
gauge theories from the point of view of D-branes, our conjecture may have
interesting implications for string field theories.

\section*{Acknowledgments}
AD would like to thank S. Minwalla for a helpful discussion and JM
would like  to acknowledge support from the  
Albert Einstein Institute and the warm hospitality of Professor
Hermann  Nicolai. 
This work is supported in part by US DOE Grant No. DE-FG 02-91ER40685.

\newpage 
\section*{Appendix 1:}

In this appendix, we indicate briefly how the infinite chain of
constraints arises with the conventional choice of static gauge. Let
us choose
\begin{equation}
X^{0} = \tau,\qquad g_{0a} = 0\quad a=1,2
\end{equation}
In such a gauge, all the reparameterization invariance is fixed and
the BLT  action takes the form
\begin{equation}
S = \int d^{3}\xi\,L = \int d^{3}\xi\, \frac{1}
{2V}\left(\overline{g}(1+\dot{X}^{M}\dot{X}_{M}) -
\widetilde{\mathcal{F}}^{2}\right)
\end{equation}
Here, the indices $M$ take only  spatial values $1,2,\cdots ,10$. The
momenta, in this gauge, have the forms
\begin{eqnarray}
P_{M} & = & \frac{\partial L}{\partial\dot{X}^{M}} =
\frac{\overline{g}}{V} \dot{X}_{M} - \frac{1}{2V}
\widetilde{\mathcal{F}}
\varepsilon^{ab}A_{MNP}\partial_{a}X^{N}\partial_{b}X^{P}\nonumber\\
\Pi^{(U) ab} & = & \frac{\partial L}{\partial\dot{U}_{ab}} = -
\frac{3\varepsilon^{ab}}{V} \widetilde{\mathcal{F}}\nonumber\\
\Pi^{(U) 0a} & = & \frac{\partial L}{\partial\dot{U}_{0a}} \approx
0\nonumber\\
P_{V} & = & \frac{\partial L}{\partial\dot{V}} \approx 0
\end{eqnarray}
Thus, we see that the velocities can be inverted and that we have two
primary constraints as before. Let us define, as in the analysis of
section {\bf 3} (except that now indices take only spatial values),
\begin{equation}
{\mathcal{P}}_{M} = P_{M} - \frac{1}{6}\,\Pi^{(U) ab} A_{MNP}
\partial_{a}X^{N}\partial_{b}X^{P}
\end{equation}
With this, as well as our assumption of a cylindrical topology for the
open membrane, the boundary conditions take the form
\begin{equation}
\left[\left({\mathcal{P}}^{2} + \left(\frac{\overline{g}}
{V}\right)^{2}\right) \overline{g}^{a1}\partial_{a}X_{\mu} + \frac{1}
{3}\Pi^{(U) a1} A_{\mu\nu\lambda}
{\mathcal{P}}^{\nu}\partial_{a}X^{\lambda}\right]_{\sigma_{1}=0,\pi} \approx 0
\end{equation}

Thus, we can write all the primary constraints, in this gauge, to be
\begin{eqnarray}
\varphi_{1} & = & P_{V} \approx 0\nonumber\\
\varphi_{2}^{a} & = & \Pi^{(U) 0a} \approx 0\nonumber\\
\varphi_{3\mu} & = & \left(\left({\mathcal{P}}^{2} +
\left(\frac{\overline{g}}
{V}\right)^{2}\right)\overline{g}^{a1}\partial_{a}X_{\mu} + \frac{1}{3}
\Pi^{(U)
a1}A_{\mu\nu\lambda}{\mathcal{P}}^{\nu}\partial_{a}X^{\lambda}\right)
\delta(\sigma_{1}) \approx 0\\
\varphi_{4\mu} & = & \left(\left({\mathcal{P}}^{2} +
\left(\frac{\overline{g}}
{V}\right)^{2}\right)\overline{g}^{a1}\partial_{a}X_{\mu} + \frac{1}{3}
\Pi^{(U)
a1}A_{\mu\nu\lambda}{\mathcal{P}}^{nu}\partial_{a}X^{\lambda}\right)
\delta(\sigma_{1}-\pi)
\approx 0\nonumber
\end{eqnarray}
The total Hamiltonian for the system, including the primary
constraints, has the form,
\begin{eqnarray}
H & = & \frac{V}{2\overline{g}} \left({\mathcal{P}}^{2} -
\left(\frac{\overline{g}}{V}\right)^{2}\right) - \frac{V}{72}
\left(\varepsilon_{ab}\Pi^{(U) ab}\right)^{2} + 2 \Pi^{(U) ab}
\partial_{a}U_{0b}\nonumber\\
 &  & + c\varphi_{1} + k_{a}\varphi_{2}^{a} +
\lambda^{\mu}\varphi_{3\mu} + \widetilde{\lambda}^{\mu}\varphi_{4\mu}
\end{eqnarray}
So far everything seems completely parallel to the discussion in
section {\bf 4} except for the extra $\left(\frac{\overline{g}}
{V}\right)^{2}$ terms in the boundary conditions as well as in the
Hamiltonian. As we will see, these make all the difference.

The constraint analysis can be carried out now. Requiring
$\dot{\varphi}_{1}\approx 0$ leads to
\begin{eqnarray}
\lambda^{\mu} & = & 0 = \widetilde{\lambda}^{\mu}\nonumber\\
\varphi_{5} & = & \frac{1}{2\overline{g}} \left({\mathcal{P}}^{2} +
\left(\frac{\overline{g}}{V}\right)^{2}\right) - \frac{1}{72}
\left(\varepsilon_{ab}\Pi^{(U) ab}\right)^{2} \approx 0
\end{eqnarray}
As before, there is a secondary constraint, which, however,
does not correspond to the vanishing of the Hamiltonian. In fact,
there is no reason for it to, since we have already gauge fixed the
$\tau$ reparameterization invariance. Requiring
$\dot{\varphi}_{5}\approx 0$, as is easily seen, determines the
Lagrange multiplier $c$. This is already a point of departure from the earlier
analysis. (Namely, earlier, the secondary condition implied the
vanishing of the Hamiltonian which is automatically invariant under
time evolution. But in the present case, the new constraint does not
correspond to the vanishing of the Hamiltonian and, therefore, leads
to  a nontrivial relation, namely it determines the Lagrange
multiplier $c$.)

Requiring $\dot{\varphi}_{2}^{a}\approx 0$ leads, as before, to the
Gauss' Law constraint
\begin{equation}
\varphi_{6}^{a} = \partial_{b}\Pi^{(U) ab} \approx 0
\end{equation}
which does not generate any further constraint. Turning now to the
boundary constraints, we recognize that, since the Lagrange multiplier $c$
is already determined and that the boundary constraints do not depend
on the field $U_{0a}$, their consistency can only lead to new
(secondary) constraints which, in turn, will lead to tertiary
constraints and so on. Explicit calculation, indeed, do verify
this. Namely, in this gauge, the boundary constraints do lead to an
infinite chain of constraints as is normally expected.

\section*{Appendix 2:}

In this appendix, we will give the explicit form of the Dirac bracket,
linear in $A_{\mu\nu\lambda}$, defined in Eq. (\ref{final}) as well as
indicate the structure of the Dirac bracket up to quadratic terms in
$A_{\mu\nu\lambda}$.

Let us recall that (see Eq. (\ref{inv1}))
\begin{eqnarray}
\left(\beta^{(0)}\right)^{\mu\nu} & = &
\left.\left(c^{-1}\right)^{\mu\nu}\right|_{A_{\mu\nu\lambda}=0} = -
\left((F+G)^{-1}\right)^{\mu\nu}\nonumber\\
\left(\gamma^{0}\right)^{\mu\nu} & = &
\left.\left(b^{-1}\right)^{\mu\nu}\right|_{A_{\mu\nu\lambda}=0} =
\left((F+G)^{-1}\right)^{\mu\nu}\nonumber\\
\left(\delta^{(1)}\right)^{\mu\nu} & = & -
\left.\left(b^{-1}ac^{-1}\right)^{\mu\nu}\right| =
\left((F+G)^{-1}(V+W)(F+G)^{-1}\right)^{\mu\nu}\label{a21}
\end{eqnarray}
The Dirac bracket between the coordinates, linear in
$A_{\mu\nu\lambda}$ (see Eq. (\ref{final})), can now be written out
explicitly as
\begin{eqnarray}
& &\left\{X_{\mu}(\sigma),X_{\nu}(\sigma')\right\}_{D}\nonumber\\
= & &
\widetilde{\Omega}_{\mu\lambda}(\sigma)
\left(\gamma^{(0)}\right)^{\lambda\rho}(\sigma,\sigma') S_{\rho\nu}(\sigma')
- \partial_{a}\left(\widetilde{\Gamma}_{\mu\lambda}^{a}(\sigma)
\left(\gamma^{(0)}\right)^{\lambda\rho}(\sigma,\sigma')
S_{\rho\nu}(\sigma')\right)\nonumber\\
 & - &
\widetilde{S}_{\mu\lambda}(\sigma)\left(\beta^{(0)}\right)^{\lambda\rho}(\sigma,\sigma')
\Omega_{\rho\nu}(\sigma') +
\partial_{a}^{\prime}\left(\widetilde{S}_{\mu\lambda}(\sigma)
\left(\beta^{(0)}\right)^{\lambda\rho}(\sigma,\sigma')\Gamma_{\rho\nu}(\sigma')\right)\nonumber\\
 & + &
\widetilde{\Omega}_{\mu\lambda}(\sigma)\left(\delta^{(1)}\right)^{\lambda\rho}(\sigma,\sigma')
\Omega_{\rho\nu}(\sigma') -
\partial_{a}\left(\widetilde{\Gamma}_{\mu\lambda}^{a}(\sigma)
\left(\delta^{(1)}\right)^{\lambda\rho}(\sigma,\sigma')
\Omega_{\rho\nu}(\sigma')\right)\label{DBlin}\\
 & - &
\partial_{a}^{\prime}\left(\widetilde{\Omega}_{\mu\lambda}(\sigma)\left(\delta^{(1)}\right)^{\lambda\rho}(\sigma,\sigma')
\Gamma_{\rho\nu}^{a}(\sigma')\right) +
\partial_{a}\partial_{b}^{\prime}\left(\widetilde{\Gamma}_{\mu\lambda}^{a}(\sigma)
\left(\delta^{(1)}\right)^{\lambda\rho}(\sigma,\sigma')
\Gamma_{\rho\nu}^{b}(\sigma')\right)\nonumber
\end{eqnarray}
We note here that the Dirac bracket can be specified completely in
terms of the coordinates and the $A_{\mu\nu\lambda}$ field with the
identification (see Eqs. (\ref{30}) and (\ref{33}))
\begin{equation}
\Pi^{(U) ab} = 3 \varepsilon^{ab}
\end{equation}

Let us next indicate briefly the structure of the Dirac bracket between
coordinates up to quadratic order in the $A_{\mu\nu\lambda}$
fields. From Eq. (\ref{DB}) and from the structure of various
quantities in there, we see that up to quadratic order in
$A_{\mu\nu\lambda}$, the Dirac bracket will contain, in addition to
the terms on the right hand side in Eq. (\ref{DBlin}), terms which are
quadratic in $A_{\mu\nu\lambda}$. These can be written symbolically as
\begin{eqnarray}
\left\{X_{\mu},X_{\nu}\right\}^{(2)}_{DB} & = & \left(\widetilde{S}\alpha^{(0)}S +
\widetilde{U}\gamma^{(0)}S + \widetilde{S}\beta^{(0)}U\right)\nonumber\\
 &  & + \left(\widetilde{T}\gamma^{(1)}S + \widetilde{S}\beta^{(1)}T +
\widetilde{T}\delta^{(1)}U +
\widetilde{U}\delta^{(1)}T\right) + \widetilde{T}\delta^{(2)}T\label{DBqua}
\end{eqnarray}
To determine this, let us note that we can decompose and write the
elements  of the matrix of constraints as a series of terms containing
different powers of $A_{\mu\nu\lambda}$. Namely, let us write
\begin{eqnarray}
a & = & a^{(1)} + a^{(2)}\nonumber\\
b & = & b^{(0)} + b^{(1)} + b^{(2)}\nonumber\\
c & = & c^{(0)} + c^{(1)} + c^{(2)}\nonumber\\
d & = & d^{(0)} + d^{(1)} + d^{(2)}
\end{eqnarray}
It is important to recognize that, since the constraints are at best
linear in $A_{\mu\nu\lambda}$, the elements of the constraint matrix
can at most have quadratic dependence on $A_{\mu\nu\lambda}$.

With this, we can determine the elements of the inverse matrix
perturbatively. At the zeroth order, they take the forms
\begin{eqnarray}
\alpha^{(0)} & = & -
\left(c^{(0)}\right)^{-1}d^{(0)}\left(b^{(0)}\right)^{-1}\nonumber\\
\beta^{(0)} & = & \left(c^{(0)}\right)^{-1}\nonumber\\
\gamma^{(0)} & = & \left(b^{(0)}\right)^{-1}\nonumber\\
\delta^{(0)} & = & 0
\end{eqnarray}
At linear order, they are determined to be
\begin{eqnarray}
\alpha^{(1)} & = &
-\alpha^{(0)}\left(a^{(1)}-b^{(0)}\left(d^{(0)}\right)^{-1}c^{(1)} -
b^{(1)}\left(d^{(0)}\right)^{-1}c^{(0)}\right.\nonumber\\
 &  & \qquad\quad \left. + 
b^{(0)}\left(d^{(0)}\right)^{-1}d^{(1)}\left(d^{(0)}\right)^{-1}c^{(0)}\right)\alpha^{(0)}\nonumber\\
\beta^{(1)} & = & - \beta^{(0)}\left(c^{(1)} -
d^{(0)}\left(b^{(0)}\right)^{-1}a^{(1)}\right)\beta^{(0)}\nonumber\\
\gamma^{(1)} & = & -\gamma^{(0)}\left(b^{(1)} -
a^{(1)}\left(c^{(0)}\right)^{-1}d^{(0)}\right)\gamma^{(0)}\nonumber\\
\delta^{(1)} & = & -
\left(b^{(0)}\right)^{-1}a^{(1)}\left(c^{(0)}\right)^{-1}
\end{eqnarray}
And, finally, the last term that is needed for the
Dirac bracket at the quadratic order (\ref{DBqua}) is determined to be
\begin{eqnarray}
\delta^{(2)}  & = &
\left(b^{(0)}\right)^{-1}a^{(1)}\left(c^{(0)}\right)^{-1}\left(c^{(1)}
-
d^{(0)}\left(b^{(0)}\right)^{-1}a^{(1)}\right)\left(c^{(0)}\right)^{-1}
\nonumber\\
 &  & \quad  -
\left(b^{(0)}\right)^{-1}\left(a^{(2)} -
b^{(1)}\left(b^{(0)}\right)^{-1}a^{(1)}\right)\left(c^{(0)}\right)^{-1}
\end{eqnarray}
Substituting these into Eq. (\ref{DBqua}) determines explicitly the
Dirac bracket for the coordinates that is quadratic in the
$A_{\mu\nu\lambda}$ fields. In fact, this iterative procedure can be
carried out to any order in the field $A_{\mu\nu\lambda}$

\end{document}